%Paper: hep-th/9301075
%From: Pierre Mathieu <pmathieu@phy.ulaval.ca>
%Date: Mon, 18 Jan 93 14:07:22 EST

%===========================================================================
% Berenstein-Zelevinsky triangles, elementary couplings and fusion rules.
%
% by  L. Begin, A.N. Kirillov, P. Mathieu and M.A. Walton
%
% "harvmac" format
%
%
\input harvmac.tex
%===========================================================================
% MACROS

\font\smallcapfont=cmr9
\def\sc#1{{\smallcapfont\uppercase{#1}}}
\def\frac#1#2{{\textstyle{#1\over #2}}}

% standard tableaux
\def\b#1{\kern-0.25pt\vbox{\hrule height 0.2pt\hbox{\vrule
width 0.2pt \kern2pt\vbox{\kern2pt \hbox{#1}\kern2pt}\kern2pt\vrule
width 0.2pt}\hrule height 0.2pt}}

\def\STrow#1{\hbox{#1}\kern-1.35pt}

% triangles pour su(3)
\def\tri#1#2#3#4#5#6#7#8#9{\matrix{#1\cr
	#2\quad#3\cr #4~\qquad #5\cr #6\quad #7\quad#8\quad#9\cr}}

% triangles pour su(4)
%\def\fbz#1#2#3#4#5#6#7#8#9#10#11#12#13#14#15#16#17#18

%{\matrix{&&&&&#1&&&&&\cr
%	&&&&#2&&#3&&&&\cr
% &&&#4&&&&#5&&&\cr
%&&#6&&#7&&#8&&#9&&\cr
%&#10&&&&#11&&&&#12&\cr
%#13&&#14&&#15&&#16&&#17&&#18 \cr}}

% hexagones
\def\hex#1#2#3#4#5#6{\matrix{	#1\quad#2\cr #3~\qquad #4\cr #5\quad
#6\cr}}

\def\text#1{\quad\hbox{#1}\quad}

\def\su{\widehat{su}}
\def\Nb{N_{\lambda\mu\nu}}

% Equations (overrides harvmac's equation macros)
\newcount\eqnum
\eqnum=0
\def\eq{\global\advance\meqno by1 \eqno(\secsym\the\meqno)}
\def\eqlabel#1{\eq {\xdef#1{\secsym\the\meqno}}}

% References (overrides harvmac's reference macros)
\newwrite\refs
\def\startreferences{
 \immediate\openout\refs=references
 \immediate\write\refs{\baselineskip=14pt \parindent=16pt \parskip=2pt}
}
\startreferences

\refno=0
\def\aref#1{\global\advance\refno by1
 \immediate\write\refs{\noexpand\item{\the\refno.}#1\hfil\par}}
\def\ref#1{\aref{#1}\the\refno}
\def\refname#1{\xdef#1{\the\refno}}

\def\pl#1#2#3{Phys. Lett. {\bf B#1#2#3}}
\def\np#1#2#3{Nucl. Phys. {\bf B#1#2#3}}

\newcount\exno
\exno=0
\def\Ex{\global\advance\exno by1{\noindent\sl Example \the\exno:
\nobreak\par\nobreak}}

\parskip=6pt
%==========================================================================
% PAGE TITRE
\Title{\vbox{\baselineskip12pt
\hbox{LAVAL-PHY-22/93}\hbox{LETH-PHY-1/93}\hbox{hepth@xxxyyyy}}}
{\vbox {\centerline{Berenstein-Zelevinsky triangles,}
\bigskip
\centerline{elementary couplings and fusion rules}
}}

\centerline{L. B\'egin$^\natural$\foot{Work supported by NSERC
(Canada).}, A.N. Kirillov$^\flat$, P. Mathieu$^\natural$\foot{Work
supported by NSERC (Canada) and
FCAR (Qu\'ebec).} and M.A. Walton$^{\sharp 1}$} \vskip.2in
\smallskip\centerline{$^\natural$ \it D\'epartement de
Physique, Universit\'e Laval, Qu\'ebec, Canada G1K 7P4}
\centerline{$^\flat$ \it Steklov Mathematical Institute, Fontanka 27,
St. Petersburg 191011, Russia}
\centerline{$^\sharp$ \it Physics Department, University of
Lethbridge, Lethbridge (Alberta) Canada T1K 3M4}
\vskip .2in
\centerline{\bf Abstract}
\bigskip
\noindent
We present a general scheme for describing $\su(N)_k$ fusion rules in
terms of elementary couplings, using Berenstein-Zelevinsky triangles.  A
fusion coupling is characterized by its corresponding tensor product
coupling (i.e. its Berenstein-Zelevinsky triangle) and the threshold level
at which it first appears.  We show that a closed expression for this
threshold level is encoded in the Berenstein-Zelevinsky triangle
and an explicit method to calculate it is presented.  In this way a
complete solution of $\su(4)_k$ fusion rules is obtained. \Date{1/93}
%========================================================================
% CORPS DE L'ARTICLE
\newsec{Introduction}
Berenstein and Zelevinsky recently made a remarkable contribution to the
classic problem of computing $su(N)$ triple tensor product
multiplicities [\ref{A.D. Berenstein and A.Z. Zelevinsky, J. Algebraic
Combinatorics 1 (1992) 7.}\refname\BZ].  They found these multiplicities to be
identical to
the number of triangles (hereafter called BZ triangles) defined by a set of
non-negative integers satisfying certain conditions.  Each BZ triangle is
associated to a particular coupling of three integrable highest weight
representations of $su(N).$

On the other hand, every coupling of a given triple
tensor product can be decomposed into a product of elementary couplings.
This decomposition is unique once the redundancies (syzygies) are
eliminated (see for instance [\ref{J. Patera and R.T. Sharp,
in {\it Lecture Notes in Physics}, vol. 84, Springer Verlag,
New York 1979.}\refname\PS] and references therein).  Here we show that the
BZ triangles provide a powerful tool for the description of these
elementary couplings. In this framework, elementary couplings are put in
correspondence with basic BZ triangles.  Then, the decomposition of a
coupling into a {\it product} of elementary ones is translated into a {\it
sum} of basic triangles.  We provide a method of construction of the set of
basic BZ triangles and we observe that syzygies can be characterized in a very
simple way.

Next, BZ triangles are used to describe fusion coefficients, or, equivalently,
restricted tensor product multiplicities.  The latter refer to the truncated
tensor product for ${\cal U}_q(su(N))$ (universal enveloping algebra of
the quantum deformations of $su(N)$) when $q$ is a root of unity
[\ref{ V. Pasquier and H. Saleur, \np330
(1990) 523.},\ref{P. Furlan, A. Ganchev and V.B. Petkova,
\np343 (1990) 205},\ref{F. Goodman and H. Wenzl, Adv. Math. {\bf
82} (1990) 244.}\refname\KW].  On the other hand, fusion coefficients
correspond to the multiplicity of the scalar representation in the triple
product of three integrable representations of the Kac-Moody algebra
$\su(N)_k$ at some fixed level $k$ [\ref{E. Verlinde,
\np300 (1988) 389.}\refname\Ver,\ref{D. Gepner and E. Witten,  \np278
(1986) 493.}\refname\GW,\ref{M.A. Walton, \np340 (1990) 777; \pl241 (1990)
365; V. Ka\v c, {\it Infinite dimensional Lie algebras}, 3rd ed. (Cambridge
University Press, 1990).}\refname\W].  The link between these dual descriptions
is $q^{N+k}=1$ [\ref{L. Alvarez-Gaum\'e, C. Gomez and G. Sierra, Phys.
Lett. {\bf B} (1989) 142.}].  For definiteness, we will use the language of
fusion rules.

The physical framework for fusion rules is conformal field theory.  In
this context, fusion rules specify the conformal families (with their
multiplicies) of the various fields arising in the expansion of the operator
product of two fields in given conformal families.  Each conformal
family is characterized by its lowest state, called a primary field.
For the special case of WZNW model with spectrum generating algebra
$\hat{g}_k$, primary fields are in one-to-one correspondence with
integrable representations of $\hat{g}_k$ [\GW,\ref{V.G.Knizhnik and A.
Zamolodchikov, Nucl. Phys.  {\bf B247} (1984) 83.}\refname\KZA].

Fusion coefficients are uniquely
characterized by [\ref{C.J. Cummins, P. Mathieu and M.A. Walton, \pl254
(1991) 390; L. B\'egin, P. Mathieu and M.A. Walton, J. Phys. A: Math. Gen.
{\bf 25} (1992) 135.}\refname\CMW,\ref{A.N. Kirillov, P. Mathieu,
D. S\'en\'echal and M.A. Walton, preprint LAVAL-PHY-20/92
(LETH-PHY-2/92), 8/92, to appear in Nucl. Phys. {\bf B}; preprint
LETH-PHY-9/92 (LAVAL-PHY-23/92), 9/92, contributed to the
proceedings of the XIXth International Colloquium on Group Theoretical Methods
in Physics, Salamanca, Spain, 29/6-4/7,1992. }\refname\KMSW]

\line{\quad (1) the corresponding tensor product coefficients; \hfill}

\line{\quad (2) the set of minimum levels $\{k_0^{(i)}\}$, at which the
various couplings,\hfill}
\line{\qquad\quad labelled by $(i)$, will first appear. \hfill}
In other words, every fusion coupling can be fully characterized by a
BZ triangle and a threshold level $k_0^{(i)}$.  Here we argue that
$k_0^{(i)}$ is encoded in the data of the corresponding BZ triangle.

Our results on fusion rules are presented as a set of
observations and conjectures.  They are illustrated by two examples:
$\su(3)_k$ and $\su(4)_k$.
%=========================================================================
\newsec{BZ triangles.}
An $su(3)$ BZ triangle, describing a particular coupling associated to the
triple product $\lambda\otimes\mu\otimes\nu$, is a triangular arrangement
of nine non-negative integers:
$$\matrix{m_{13}\cr
	n_{12}~~\quad l_{23}\cr
m_{23}~\quad\qquad ~~m_{12}\cr
 n_{13}~\quad l_{12} \qquad n_{23} \quad~ l_{13} \cr
}\eqlabel\trbz$$
These integers are related to the Dynkin labels of the three integrable
highest weights by $$\matrix{m_{13}+n_{12}=\lambda_1 \quad n_{13}+l_{12}=\mu_1
\quad l_{13}+m_{12}=\nu_1\cr
 m_{23}+n_{13}=\lambda_2 \quad n_{23}+l_{13}=\mu_2 \quad
l_{23}+m_{13}=\nu_2\cr}\eqlabel\wee$$ and they further satisfy the so-called
hexagon conditions $$\matrix{n_{12}+m_{23} = n_{23}+m_{12}\cr l_{12}+m_{23} =
l_{23}+m_{12}\cr l_{12}+n_{23} = l_{23}+n_{12}\cr }\eqlabel\ghr$$
These last conditions mean that the length of opposite sides in the
hexagon are equal, the length of a segment being defined as the sum of its
two vertices.  An $su(3)$ BZ triangle is thus composed of one hexagon and
three corner points.

Each pair of indices $ij,\ i<j,$ on the labels of the triangle indicates
association with a positive root of $su(3).$ If $e_i$ are orthonormal vectors
in ${\bf R}^N,$ then the positive roots of $su(N)$ can be represented in
the form $e_i-e_j,\ 1\leq i<j\leq N.$ The triangle encodes three sums of
positive roots:
$$\matrix{\mu +\nu -C\lambda\ =\ \sum_{i<j}\ l_{ij} (e_i-e_j)\ \ ,\cr
\nu +\lambda -C\mu\ =\ \sum_{i<j}\ m_{ij} (e_i-e_j)\ \ ,\cr
\lambda +\mu -C\nu\ =\ \sum_{i<j}\ n_{ij} (e_i-e_j)\ \ ,\cr}
\eqlabel\expansions$$
where $C\lambda$ is the weight contragredient (charge conjugate) to the weight
$\lambda.$ The hexagon relations (\ghr) can then be seen as consistency
conditions for these three expansions.

For $su(4)$ the BZ triangle is defined in a similar way, in terms of
eighteen non negative integers:
$$\matrix{m_{14}\cr
	n_{12}~~\quad l_{34}\cr
m_{24}~\qquad\qquad ~~m_{13}\cr
n_{13}\qquad l_{23}\qquad n_{23} \qquad l_{24}\cr
m_{34}\qquad\qquad\quad m_{23}\qquad\qquad\quad m_{12} \cr
n_{14}\qquad~~ l_{12}\qquad~ n_{24}\quad~\quad l_{13}\quad~~~ n_{34}\qquad
l_{14} \cr  }\eqlabel\fobz$$

\noindent related to the Dynkin labels by
$$\matrix{\matrix{m_{14}+n_{12} =\lambda_1\cr m_{24}+n_{13} =\lambda_2\cr
m_{34}+n_{14} = \lambda_3\cr } \quad \matrix{n_{14}+l_{12} =\mu_1\cr
n_{24}+l_{13} =\mu_2 \cr n_{34}+l_{14} =\mu_3~~\cr }
\quad \matrix{l_{14}+m_{12}
=\nu_1\cr
l_{24}+m_{13} =\nu_2\cr ~~l_{34}+m_{14} =\nu_3
\cr } \cr}\eqlabel\mma$$
Furthermore, the $su(4)$ BZ triangle contains three hexagons:
$$\matrix{\eqalign{n_{12}+m_{24} &=m_{13}+n_{23}\ \cr n_{12}+l_{34}
&=l_{23}+n_{23}\ \cr m_{24}+l_{23} &=l_{34}+m_{13}\ \cr\cr }
\eqalign{n_{13}+l_{23} &=l_{12}+n_{24}~~\cr n_{13}+m_{34} &=n_{24}+m_{23}\cr
{}~m_{34}+l_{12} &= l_{23}+m_{23}\cr\cr }
\eqalign{\ l_{24}+n_{23} &=l_{13}+n_{34}~~\cr\ n_{23}+m_{23} &=m_{12}+n_{34}
\cr\ ~~l_{13}+m_{23} &=l_{24}+m_{12} \cr\cr }}\eqlabel\mmb$$

The $su(N)$ generalization is obvious; the triangle is built out of
$(N-1)(N-2)/2$ hexagons and three corner points.

The relation between BZ triangles and tensor product multiplicities is the
following:  For a fixed triplet $(\lambda,\mu,\nu)$ of highest weights of
integrable representations, the number of possible BZ triangles gives
the multiplicity of the scalar representation in the triple tensor product
$\lambda\otimes\mu\otimes\nu$ [\BZ].

\Ex{}
The five BZ triangles associated to the $su(4)$ tensor product
$(1,2,1)^{\otimes 3}$ are:
$$\matrix{\matrix{1\cr
	0~~\quad 0\cr
 1~\quad\quad ~~1\cr
 1~\quad 0\quad 0 \quad ~1\cr
0\qquad\quad 0\qquad\quad 0 \cr
1~\quad 0~\quad 1~\quad 1~\quad 0~\quad 1 \cr
}
&\matrix{0\cr
	1~~\quad 1\cr
 1~\quad\quad ~~1\cr
 1~\quad 1\quad 1 \quad ~1\cr
1\qquad\quad 1\qquad\quad 1 \cr
0~\quad 1~\quad 1~\quad 1~\quad 1~\quad 0 \cr
}
&\matrix{1\cr
	0~~\quad 0\cr
 0~\quad\quad ~~0\cr
 2~\quad 0\quad 0 \quad ~2\cr
1\qquad\quad 2\qquad\quad 1 \cr
0~\quad 1~\quad 1~\quad 1~\quad 1~\quad 0 \cr
}}$$
$$\matrix{\matrix{0\cr
	1~~\quad 1\cr
 1~\quad\quad ~~2\cr
 1~\quad 2\quad 0 \quad ~0\cr
1\qquad\quad 0\qquad\quad 0 \cr
0~\quad 1~\quad 2~\quad 0~\quad 0~\quad 1 \cr }
&\matrix{0\cr
	1~~\quad 1\cr
 2~\quad\quad ~~1\cr
 0~\quad 0\quad 2 \quad ~1\cr
0\qquad\quad 0\qquad\quad 1 \cr
1~\quad 0~\quad 0~\quad 2~\quad 1~\quad 0 \cr
}}\eqlabel\abc$$

Finding all possible BZ triangles for fixed highest weights
$(\lambda,\mu,\nu)$ might appear difficult at first sight. However,
once one is found, all other ones are obtained by addition or substraction
of few building block triangles incorporating negative entries.
This fact was exploited for $su(3)$
in [\ref{L. B\'egin, P. Mathieu and M.A. Walton, Mod. Phys. Lett. A, Vol. 7
, No. 35 (1992) 3255.}\refname\LB]. For $su(4)$, one can show that if
$\Delta$ is a BZ triangle for fixed $(\lambda,\mu,\nu),$
then all others are of the form
$\Delta + c_1\delta_1 +c_2\delta_2 + c_3\delta_3,$
where the $\delta_i$ are, respectively,
$$\matrix{\matrix{\bar1\cr
	1~~\quad 1\cr
 1~\quad\quad ~~1\cr
 \bar1~\quad 1\quad 1 \quad ~\bar1\cr
0\qquad\quad \bar1\qquad\quad 0 \cr
0~\quad 0~\quad 0~\quad 0~\quad 0~\quad 0 \cr
}
&\matrix{0\cr
	0~~\quad 0\cr
 \bar1~\quad\quad ~~0\cr
 1~\quad 1\quad \bar1 \quad ~0\cr
1\qquad\quad 1\qquad\quad 0 \cr
\bar1~\quad 1~\quad 1~\quad \bar1~\quad 0~\quad 0 \cr
}
&\matrix{0\cr
	0~~\quad 0\cr
 0~\quad\quad ~~\bar1\cr
 0~\quad \bar1\quad 1 \quad ~1\cr
0\qquad\quad 1\qquad\quad 1 \cr
0~\quad 0~\quad \bar1~\quad 1~\quad 1~\quad \bar1 \cr
}}\eqlabel\rnt$$
($\bar1:=-1$)\ and the $c_i$ are integers. For example, the five
BZ triangles of (\abc) can be expressed as
$\{ \Delta-(\delta_1+\delta_2+\delta_3), \Delta, \Delta-\delta_1,
\Delta-\delta_3, \Delta-\delta_2
\},$ respectively.

\newsec{Basis for BZ triangles.}
{}From the linearity of the conditions defining the BZ triangles, it is clear
that every BZ triangle can be decomposed into a sum of basic triangles whose
entries take values in the set $\{0,1\}$.  We give a construction of a
minimal set of basic triangles.

Three basic triangles are easily described: they have $0'$s everywhere and
a single 1 at one of the three corners.  The other basic triangles have 0
at each corner and some 1's distributed among the hexagons such that each
hexagon contains at most three 1's (actually, it can have zero, two or
three 1's), with at least one hexagon being non empty.  Every inequivalent
irreducible distribution produces an independent basic triangle.

\Ex{}
The $su(3)$ basic triangles are
$$\matrix{E_1=(0,0)(1,0)(0,1)\cr~\cr\tri001000100}\qquad
\matrix{E_3=(1,0)(0,1)(0,0)\cr~\cr\tri010000010}\qquad
\matrix{E_5=(0,1)(0,0)(1,0)\cr~\cr\tri000110000}$$
$$\matrix{E_2=(1,0)(0,0)(0,1)\cr~\cr\tri100000000}\qquad
\matrix{E_4=(0,1)(1,0)(0,0)\cr~\cr\tri000001000}\qquad
\matrix{E_6=(0,0)(0,1)(1,0)\cr~\cr\tri000000001} $$
$$\matrix{E_7=(0,1)(0,1)(0,1)\cr~\cr\tri001100010}\qquad
\matrix{E_8=(1,0)(1,0)(1,0)\cr~\cr\tri010010100} $$
One can introduce a compact notation to specify the hexagon content as
well as the exact position of the 1's inside the hexagon:
$$2=\matrix{\hex010010}\qquad
2'=\matrix{\hex100001}\qquad
2''=\matrix{\hex001100} \eqlabel\abce$$

$$3=\matrix{\hex011001}\qquad
3'=\matrix{\hex100110}\qquad \eqlabel\abcf$$

\Ex{}
With this notation and the hexagon ordering $\matrix{1 \cr 2 \quad 3 \cr}$,
one can write the full set of basic triangles (that is the full set of
elementary couplings) for $su(4)$: (compare with [\ref{C.J. Cummins, M.
Couture and R.T. Sharp, J. Phys. A: Math. Gen. {\bf 23} (1990)
1929.},\ref{R.T. Sharp and D. Lee, Revista Mexicana de Fisica {\bf 20}
(1971) 203-215.}]) $$\matrix{\eqalign{A_1&=\underline{0}00\cr
A_2&=0\underline{0}0\cr A_3&=00\underline{0}\cr B_3&=2''00 \cr B_2&=02'0
\cr B_1&=002	\cr}
&\eqalign{C_1&=220	\cr
C_2&=2'02'	\cr
C_3&=02''2'' \cr
D_1&=302' \cr
D_2&=3'20  \cr
D_3&=230  \cr}
&\eqalign{D_1'&=03'2''  \cr
D_2'&=02''3  \cr
D_3'&=2'03'  \cr
E_1&=3'30  \cr
E_2&=303'   \cr
E_3&=03'3    \cr}} \eqlabel\mmmh$$
The underlined zero means that one places a 1 at the corner adjacent to
the corresponding hexagon.  For instance

$$0\underline{0}0=\matrix{0\cr
	0~~\quad 0\cr
 0~\quad\quad ~~0\cr
 0~\quad 0\quad 0 \quad ~0\cr
0\qquad\quad 0\qquad\quad 0 \cr
1~\quad 0~\quad 0~\quad 0~\quad 0~\quad 0 \cr
}\eqlabel\abcg$$
corresponds to the coupling $(0,0,1)(1,0,0)(0,0,0)$.  Similarly
$$03'2''={\matrix{0\cr
	0~~\quad 0\cr
 0~\quad\quad ~~0\cr
 1~\quad 0\quad 0 \quad ~0\cr
0\qquad\quad 1\qquad\quad 1 \cr
0~\quad 1~\quad 0~\quad 0~\quad 0~\quad 0 \cr
}}\eqlabel\abch$$
describes the coupling $(0,1,0)(1,0,0)(1,0,0)$. To illustrate the
irreducible character of the elementary couplings, notice that the
following is a consistent way of distributing the 1's in the three hexagons:
$$3'32={\matrix{0\cr
	1~~\quad 0\cr
 0~\quad\quad ~~1\cr
 0~\quad 1\quad 0 \quad ~1\cr
1\qquad\quad 0\qquad\quad 0 \cr
0~\quad 0~\quad 1~\quad 1~\quad 0~\quad 0 \cr
}}\eqlabel\abci$$
but it is reducible, e.g.: $3'32=3'30+002$.

%=========================================================================
\newsec{Characterization of the syzygies}
The decomposition of a general BZ triangle into a sum of basic triangles
is unique only after all the redundancies are eliminated.  From the point
of view of BZ triangles, syzygies can be characterized by a number of adjacent
hexagons. Explicitly, the sources of all redundancies for $su(N)$
(at least for
$su(N<6)$) are the
following two non-unique decompositions:
$$6=\hex111111 \quad\qquad
6=3+3'=2+2'+2'' \eqlabel\mmc$$ and, with $5=2+3,5'=2+3'$, e.g.
$$55'=\matrix{
	1\quad 1\cr
 0~\quad\quad ~~1\cr
 0~\quad 2\quad 0 \quad\quad\cr
1\qquad\quad 0\qquad\quad \quad \cr
\qquad~\quad 1~\quad 1~\quad \quad~\quad \quad~\quad \quad \cr
}\eqlabel\mmdas$$
$$ 55'=22+3'3=23+3'2 \eqlabel\mmd$$

\Ex{}
For $su(3)$ there is thus only one syzygy, namely $(\mmc)$.  In terms of
elementary couplings, it reads $E_7 E_8=E_1 E_3 E_5$.  An example of the
second type for $su(4)$ is $C_1 E_1=D_3 D_2$.  The full list contains 15
syzygies, among which only three are of the second type:
$$\matrix{\eqalign{C_1 E_1=D_2 D_3 \cr
C_2 E_2=D_1 D_3'  \cr
C_3 E_3=D_2' D_1'  \cr
B_3 C_1 C_2=D_1 D_2  \cr
B_2 C_1 C_3=D_3 D_1' \cr}
&\eqalign{B_1 C_2 C_3=D_2' D_3' \cr
E_1 E_2=B_3 D_3 D_3'  \cr
E_1 E_3=B_2 D_2 D_2'  \cr
E_2 E_3=B_1 D_1 D_1'   \cr
D_1 E_1=B_3 C_2 D_3    \cr}
&\eqalign{D_2 E_2=B_3 C_1 D_3' \cr
D_3 E_3=B_2 C_1 D_2'  \cr
D_1' E_1=B_2 C_3 D_2  \cr
D_2' E_2=B_1 C_3 D_1  \cr
D_3' E_3=B_1 C_2 D_1'  \cr}} \eqlabel\mmmk$$

%========================================================================
\newsec{Fusion coefficients and threshold levels.}
Recall that a fusion triple product refers to a product (denoted by
$\times$) of three integrable highest weight representations of a Kac-Moody
algebra $\hat{g}$ at some positive integer level $k$.  Such
representations are characterized by a highest weight whose Dynkin labels
are integers and satisfy the inequality $(\lambda,\theta)\leq k$, where
$\theta$ is the longest root.  To $\lambda$ we then associate the affine
weight $\hat{\lambda}$ obtained by the addition of a zeroth Dynkin label
$\lambda_0=(\lambda,\theta)-k$.  Hence all the Dynkin labels of the highest
weight of an integrable representation, including the zeroth one, must
be non-negative integers.  The fusion coefficients
$N^{(k)}_{\hat{\lambda}\hat{\mu}\hat{\nu}}$ gives the multiplicity of the
scalar representation in the triple product
$\hat{\lambda}\times\hat{\mu}\times\hat{\nu}$.

The method of generating functions for fusion rules [\CMW] as
well as the depth rule [\GW,\KMSW], suggest
that an efficient description of fusion coefficients consists in specifying
the minimum level at which every coupling is first allowed (see also
[\LB]).  We
label the couplings by an index $(i)$ running from 1 to $N_{\lambda\mu\nu}$,
and
denote the threshold level associated to the coupling $(i)$ as $k_0^{(i)}$.
We
will assume that the couplings are ordered such that $k_0^{(i)}\leq
k_0^{(i+1)}$.  Then the relation between fusion coefficients and the data
$N_{\lambda\mu\nu}$ and $\{k_0^{(i)}\}$ is
$$N^{(k)}_{\hat{\lambda}\hat{\mu}\hat{\nu}} = \left\{ \eqalign{\max&(i)
\text{such that} k \geq {k_{0}}^{(i)} \text{and} \Nb\neq 0 \cr    0 & \quad~
\text{if} k<{k_{0}}^{(1)} \text{or} \Nb=0. \cr} \right.\eqlabel\abckp$$

For later reference, we recall how the group of outer automorphisms
${\cal O}(\hat{g})$ acts on fusion coefficients [\ref{J. Fuchs and D.
Gepner, \np294 (1987) 30; J. Fuchs, Nucl. Phys. B (Proc. Suppl.) {\bf 6}
(1989) 157.}]: $$N^{(k)}_{\hat{\lambda}\hat{\mu}\hat{\nu}}= N^{(k)}_{A
\hat{\lambda}, A' \hat{\mu}, A''\hat{\nu}} \quad \text{if} A A'
A''=1\eqlabel\abclk$$ $A,A'\text{and}A''$ are three elements of ${\cal
O}(\hat{g})$.  For $\su(N)$, any element of the outer automorphism group can
be written as a power of $a$, defined as $$a \hat{\lambda}=a[\lambda_0,...,
\lambda_{n-1}]=[\lambda_{n-1},\lambda_0,
\lambda_1,...,\lambda_{n-2}]\eqlabel\ajh$$ (We use square brackets when the
zeroth Dynkin label is included and parentheses otherwise.)

%=========================================================================
\newsec{Elementary couplings for fusion rules.}
An arbitrary fusion coupling
$(\hat{\lambda}\times\hat{\mu}\times\hat{\nu})^{(i)}$ may be expressed as a
product of the elementary fusion couplings ${\cal F}_j$:
$$(\hat{\lambda}\times\hat{\mu}\times\hat{\nu})^{(i)}=\prod_j {\cal
F}_j^{f_j}\ \ .\eqlabel\abcl$$
Before the syzygies are taken into account, this decomposition is not
unique. Each elementary coupling ${\cal F}_j$ has a threshold level
$k_0({\cal F}_j).$ In reference [\CMW] it was conjectured that there exists a
choice of elementary couplings, and a way to eliminate their syzygies, such
that the
decomposition $(\abcl)$ is unique, and
$$k_0^{(i)}={\sum}_j f_j k_0({\cal F}_j)\ \ \ . \eqlabel\abcv$$ Here we
assume the validity of this conjecture.

Let us now describe the set $\{{\cal F}_j\}$.  First, it contains all
elementary couplings for tensor products, which we denote by $\{{\cal
E}_j\}$.  Their $k_0$ value is easily computed using tensor product
multiplicities and the affine Weyl group [\W]. The results always
turn out to be given by:
$$\quad k_0({\cal E}_j)=[{|{\cal E}_j| \over
2}]\eqlabel\abck$$
\vskip.2cm
\noindent where $|{\cal E}_j|$ is the sum of the Dynkin
labels of the three weights in the coupling ${\cal E}_j$,
and [ ] stands for the integer part.

\Ex{}
Directly from eq. (\abck) one has:
$$\eqalign{\su(3):& \quad k_0(E_i)=1 \cr
 \su(4):& \quad
k_0(A_i)=k_0(B_i)=k_0(C_i)=k_0(D_i)=k_0(D'_i)=1;~k_0(E_i)=2
\cr}\eqlabel\abcp$$

Now consider the affine extension of the couplings
$\{{\cal E}_j\}$ at level $k_0({\cal E}_j)$, which we denote as
$\widehat{{\cal E}_j}$. At least for $su(N\leq 5),$ all possible
actions of the
outer automorphism group on these $\widehat{{\cal E}_j}$'s produce
the remaining
elements in the set $\{{\cal F}_j\}$.  Write the finite form of these extra
elements as ${\cal A}{\cal E}_j$.  Clearly one has $k_0({\cal A}{\cal
E}_j)=k_0({\cal E}_j)$. The augmented set is not minimal, however, since many
couplings of the form ${\cal A}{\cal E}_j$ can be decomposed into products of
those in the set $\{{\cal E}_j\}$.

For $su(N\leq 5)$, the minimal set of fusion elementary couplings includes
$\{{\cal E}_j\}$, the set of tensor product elementary couplings (with
$k_0({\cal E}_j)$ given by (\abck)). All other fusion elementary couplings
may be obtained from $(1,0,...,0,1)$ $(1,0,...,0,1)$ $(1,0,...,0,1)$
(with $k_0=2$) by the action of the outer
automorphism group, with one weight fixed.  For $su(2)$ and $su(3),$
the set $\{{\cal E}_j\}$ is sufficient.

\Ex{}
For $su(3)$, the coupling $(1,1)(1,1)(1,1)$ at level 2 can be
obtained from $E_7 E_8$.  For $su(4)$, among the set $\{{\cal E}_j\}$,
there is $(0,1,0)(0,1,0)(1,0,1)$ which by eq.(\abck) has
$k_0=2$.  At level $k=2$ one then has the elementary fusion coupling
$[1,0,1,0]\times [1,0,1,0]\times [0,1,0,1]$.  Another
allowed fusion coupling is $a^3[1,0,1,0]\times a[1,0,1,0]\times [0,1,0,1]$
whose finite part yields the coupling $(1,0,1)(1,0,1)(1,0,1)$.  This last
coupling has then $k_0=2$.  The other possible actions of the outer
automorphism group do not produce new independent elementary fusion
couplings.  Now $(1,0,1)^{\otimes 3}$ has two decompositions, $A_1 A_2 A_3$
and $C_1 C_2 C_3$, which both give $k_0=3$ (since $k_0(A_i)=k_0(C_i)=1$).
The idea is thus to forbid $C_1 C_2 C_3$,
and replace it by a new elementary coupling with $k_0=2$,
denoted say by $F$.

%=========================================================================
\newsec{Threshold level of arbitrary couplings.}
We now present our main conjecture, which leads to an explicit expression
for $k_0^{(i)}$ in terms of the entries in the BZ triangle associated to
the coupling $(i)$.(Given the existence of fusion elementary couplings,
the following conjecture can be viewed as a sharpened version of
the conjecture in [\CMW] mentioned in the previous section.)

$\hfill\break Conjecture:$ \ \ Before eliminating syzygies,
$$k_0(\lambda\times\mu\times\nu)^{(i)}={\rm min}(\sum f_j k_0({\cal
F}_j))\ \ ,\hfill\eqlabel\co$$
 $\hfill\break$where the minimum is taken over all
possible decompositions $\prod_j {\cal F}_j^{f_j}$.

This conjecture can also be rephrased in terms of the
specification of a set of forbidden couplings (eliminating then the
syzygies), such that the $k_0$ value of a coupling can be obtained by
the sum of its component elementary couplings.  With redundancies
of the form
$\prod_i {\cal F}_i^{e_i}=\prod_i {\cal F}'^{e_i'}_i$ one eliminates the
product of couplings with highest values of $k_0$.  In other words, if
$\sum e_i' k_0({\cal F}'_i)\geq\sum e_i k_0({\cal F}_i)$, one eliminates the
product $\prod_i {\cal F}_i'^{e_i'}$. When they have the
same value of $k_0$, which product is forbidden is immaterial.

\Ex{}
For $su(3)$, the only redundancy is $E_7 E_8=E_1 E_3 E_5$ with left
hand side having $k_0=2,$ and right hand side $k_0=3.$  Hence one should
eliminate $E_1 E_3 E_5$.  For $su(4)$, one eliminates all the products on the
l.h.s of the first nine syzygies given in (\mmmk); for the remaining six
(and as far as the calculation of $k_0$ is concerned), the
choice is arbitrary.

%=========================================================================
\newsec{Threshold level in terms of BZ triangle data:$~\su(3)_k$ and
$\su(4)_k$.}

The value of $k_0$ associated to the $su(3)$ BZ triangle (\trbz) as
calculated from (\co), is
$$k_0=\max\{m_{13}+\mu_1+\mu_2,n_{13}+\nu_1+\nu_2,l_{13}+\lambda_1+\lambda_2\}
\eqlabel\mme$$ Similarly, the threshold level of the $su(4)$ BZ
triangle (\fobz) is $$\eqalign{k_0=\max\Bigl\{&m_{14}+n_{14}+k_0(\Delta_{3}),
\cr
&l_{14}+m_{14}+k_0(\Delta_{2}),\cr
&n_{14}+l_{14}+k_0(\Delta_{1}), \cr
&l_{14}+m_{14}+n_{14}+[{\lambda_2+\mu_2+\nu_2+l_{23}+m_{23}+n_{23}+1
\over 2}] \Bigr\}\cr}\eqlabel\kli$$ where $k_0(\Delta_{i})$ refers to the
value of $k_0$ for the $su(3)$ BZ triangle circumscribing the hexagon of type
$i=\matrix{1 \cr 2 \quad 3 \cr}$ (see Example 3).  These values of $k_0$ are
computed using (\mme).

The formula for $su(3)$ was proved in [\KMSW]. The derivation of the $su(4)$
formula is a straightforward, albeit complicated, generalization of that for
$su(3).$ An alternative approach for $su(3)$ is presented in [\LB].

\Ex{}The values of $k_0$ for the five BZ triangles of example 1 are
respectively $\{6,5,5,5,5\}$.
Notice that the value of $k_0$ of two triangles is not additive.
For instance the second BZ triangle of the example 1 has $k_0=5$ while its
double $$\matrix{0\cr
	2~~\quad 2\cr
 2~\quad\quad ~~2\cr
 2~\quad 2\quad 2 \quad ~2\cr
2\qquad\quad 2\qquad\quad 2 \cr
0~\quad 2~\quad 2~\quad 2~\quad 2~\quad 0 \cr
} \eqlabel\ouib$$
has $k_0=9$.  This of course is due to the possible contractions induced
by the syzygies. For example, adding the triangle with
decomposition $C_1 C_2$
 $(k_0=2)$ to that corresponding to $C_3$ $(k_0=1)$ yields a triangle
associated to $F$, with also $k_0=2$.

\vskip2cm
\centerline{\bf Acknowledgment}
We thank R.T Sharp for useful discussions and D. S\'en\'echal for making
available to us his computer program for fusion rules.

%=========================================================================

\bigskip \hrule \bigskip \centerline{\sc{references}}
 \immediate\closeout\refs \vskip 0.5cm
  \message{References}\input references
\bye